\documentclass[prb, twocolumn, superscriptaddress, showpacs,floatfix]{revtex4}  
\usepackage{dcolumn}
\usepackage{hyperref}
\usepackage{amssymb}
\usepackage{amsmath}
\usepackage{float}
\usepackage{bbm}
\usepackage{graphicx}
\usepackage{epsfig}
\usepackage{epstopdf}
\usepackage[usenames]{color}
\begin{document}

\title{Dynamical Mean Field Theory of Nickelate Superlattices}

\author{M. J. Han} 
\altaffiliation[Current address:]{ Advanced Photon Source
  (401/B3149C), Argonne National Laboratory, Argonne, Illinois 60439,
  USA}
\affiliation{Department of Physics, Columbia University, 538 West 120$^{th}$ Street, New York, New York 10027, USA}
\author{Xin Wang}
\affiliation{Condensed Matter Theory Center, Department of Physics, University of Maryland, College Park, Maryland 20742, USA}
\author{C. A.  Marianetti}\affiliation{Department of Applied
  Physics, Columbia University, New York, New York 10027, USA}
 \author{A. J. Millis}
\affiliation{Department of Physics, Columbia University, 538 West 120$^{th}$ Street, New York, New York 10027, USA}

\date{\today }

\begin{abstract} 
Dynamical mean field methods are used to calculate the phase diagram, many-body density of states, relative orbital occupancy  and Fermi surface shape for a realistic model of $LaNiO_3$-based superlattices. The model is derived from density functional band calculations and  includes oxygen orbitals.   The combination of the on-site Hunds interaction and  charge-transfer between the transition metal and the oxygen orbitals is found to  reduce the orbital polarization far below the levels predicted either by band structure calculations or by many-body analyses of Hubbard-type models which do not explicitly include the oxygen orbitals. The findings indicate that  heterostructuring is unlikely to produce one band model physics and demonstrate the fundamental inadequacy of modeling the physics of late transition metal oxides with  Hubbard-like models. 
\end{abstract}


\maketitle

The electronic properties of transition metal oxides are of central  importance to condensed matter physics and materials science, both for their fundamental scientific interest\cite{Imada98,Nagaosa00} and for their potential for novel applications.\cite{Mannhart07} Of particular current interest are the new possibilities enabled by advances in atomic-scale  layer-by-layer growth of combinations of complex oxide materials \cite{Ahn06,Mannhart08}. Experiments report remarkable and unexpected properties including interface superconductivity,\cite{Reyren07} orbital reconstruction,\cite{Chakhalian07} high Curie-temperature magnetism \cite{Luders09} and metal-insulator transitions.\cite{Liu11,Schweritzl11}

These developments suggest that it may become possible to design materials with desired  'correlated electron' properties such as high-temperature superconductivity.  Rational materials design requires knowledge of the structure-property relation between atomic arrangement and electronic density of states. In systems such as conventional semiconductors  where the electronic structure is well described by density functional band theory this issue is well understood, so that e.g. band-gap engineering by choice of superlattice or size of quantum dot is now routine.\cite{Capasso87} However, band theory provides an incomplete description of the  relevant  electronic states in materials with strong electronic correlations,  and as a result  the understanding of  structure-property  relations in transition metal oxides is much less well developed. 

The important states  transition metal oxides are the transition metal $d$-orbitals. In the $CuO_2$-based high-$T_c$ cuprates the nominal electronic configuration of the $Cu$ is $d^9$ (one hole in the $d$-shell) and the  crystal structure breaks the rotational symmetry to the point that the only relevant $Cu$ orbital is the $d_{x^2-y^2}$.  The resulting quasi-two-dimensional $x^2-y^2$-dominated electronic structure is believed to be crucial to the high transition temperature.\cite{Anderson87,Zhang88}  In the related material $LaNiO_3$  the nominal configuration is $Ni$ $d^7$ and the  relevant orbitals are $d_{x^2-y^2}$ and $d_{3z^2-r^2}$, which transform as a doublet under crystal symmetry operations. A recent paper suggested that in a superlattice composed of alternating layers of $LaNiO_3$ and a related but insulating material such as $LaAlO_3$ the $d_{3z^2-r^2}$ orbital would be pushed so far away in energy that the Fermi surface would only have one sheet possibly leading to high temperature superconductivity.\cite{Chaloupka08} While there is to our knowledge no experimental evidence for superconductivity in nickelate superlattices,  essential aspects of correlated-electron behavior including metal-insulator transitions and magnetic behavior are intimately connected to orbital physics,\cite{Nagaosa00,Werner09} so  the issue of orbital polarization in superlattices and its relation to electronic behavior is an essential question for the subject of 'correlated electron materials by design'.

Density functional band theory calculations \cite{Hansmann09,Han10} indicate that  an $LaNiO_3/LaXO_3$ superlattice  has partial ($\sim 20-40\%$) polarization of the $Ni$ $d$ orbitals, with the precise amount (and sign) of the polarization depending on the specific choice of counter-ion $X$ in the heterostructure,\cite{Han10} but the known inadequacy of band theory for charge transfer materials such as the rare earth nickel oxides motivates an examination of the effects of correlations. In this paper we show that  strong correlation effects actually {\em decrease} the polarization. The key feature aspect of our calculation is the use  of a realistic Hamiltonian which is derived from the density functional band calculations of Ref.~\onlinecite{Han10} and in particular takes oxygen states into account explicitly. The possibility of charge transfer to the oxygen affects the results in an important way.   The model is of the general form
\begin{equation}
H=H_d+H_{hyb}+H_{ligand}
\label{Hdef}
\end{equation}
The Hamiltonian for the  correlated (``$d$'') subspace  is
\begin{equation}
H_d=\sum_{j,a,\sigma}\varepsilon_dd^\dagger_{ja\sigma}d_{ja\sigma}+\frac{U}{2}{\hat N}^j_d\left({\hat N}^j_d-1\right)+H_J
\label{Hddef}
\end{equation}
with $a=1,2$ labelling the $x^2-y^2$ and $3z^2-r^2$ d states, ${\hat N}^j_d$ the total number of d electrons on site $j$, $U$ the on-site repulsion and $H_J$ the additional 'Hunds' interactions which  give the multiplet structure at given $d$ occupation. Consistent with recent downfolding studies of the screened interaction matrix elements in transition metal oxide compounds.\cite{Aryasetiawan06,Karlsson10} we take the interaction terms to have the standard Slater-Kanamori form,\cite{Imada98} set $J=0.5eV$ and consider a range of $U$. 

$H_{hyb}$ and $H_{ligand}$ describe the $d-p$ hybridization and the ligand (oxygen) part of the band structure. They may be obtained by fitting  a tight binding model to a calculated band structure or, equivalently,  by constructing Wannier functions from states within  some appropriately chosen energy window.\cite{Anisimov91,Amadon08} The resulting tight binding model (obtained from non-spin-polarized GGA band calculations \cite{Han10}) involves $e_g$ orbitals as well as  oxygen $2p_\sigma$ orbitals with a mean oxygen energy $\varepsilon_p$, oxygen-oxygen hoppings $t_{pp}$ and $Ni$-$O$ hoppings $t_{pd}$ and is  described in detail in Ref.~\onlinecite{Han10}. The choice  $X=In$ was found to produce the largest  $d_{x^2-y^2}$ occupancy and a very  weak hybridization in the direction transverse to the superlattice plane. In this paper we  study the extreme case of a two dimensional system defined using the $Ni$ and $O$ parameters obtained from  the $LaNiO_3/LaInO_3$ superlattice studied in Ref.~\onlinecite{Han10} but with the hopping from the apical $O$ ion to the $X$ plane set to zero.   

\begin{figure}[t]
  \centering \includegraphics[width=0.85\columnwidth]{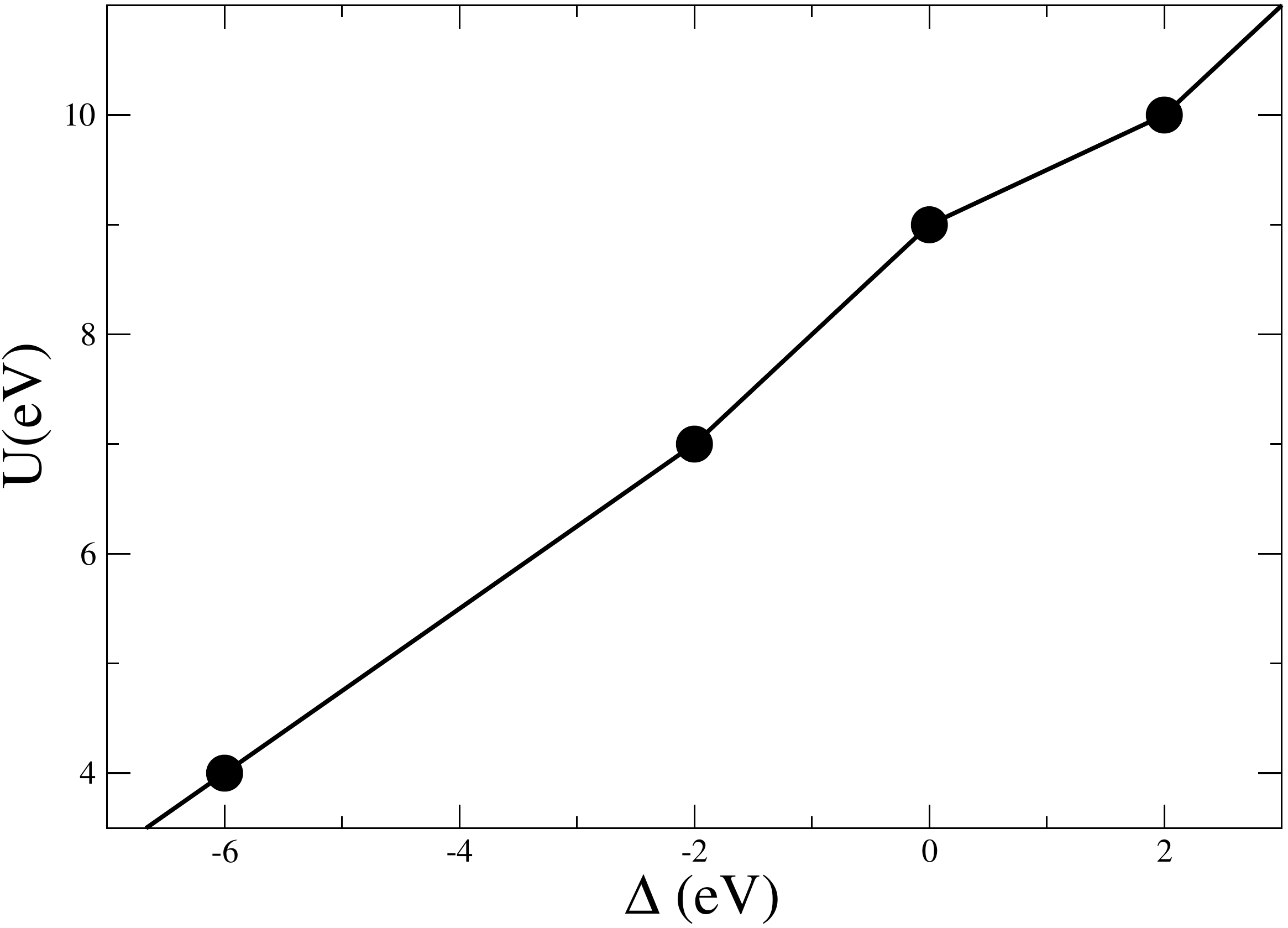}
    \caption{Metal-insulator  phase diagram of two dimensional $Ni-O$ superlattice calculated using single-site dynamical mean field theory in plane of on-site interaction strength $U$ and charge transfer energy $\Delta=\varepsilon_{p}-\varepsilon_{d}$. Plotted is $U_{c1}$,  the lower limit of stability of the insulating phase  }
   \label{phasedia}
\end{figure}

The charge transfer energy $\Delta=\varepsilon_p-\varepsilon_d$ plays a crucial role in the theory of the metal-insulator transition in oxides.\cite{Zaanen85}  $\Delta$ is renormalized from the band theory value by the double-counting correction \cite{Anisimov91} whose value is an important unsolved problem in materials theory.  Different prescriptions  have been proposed,\cite{Anisimov91,Amadon08,Karolak10} but no clear consensus has emerged.  We study a range of $\Delta$ here, corresponding to a range of double counting corrections. 

To treat the many-body physics we use the single-site dynamical mean field approximation \cite{Georges96} with the hybridization expansion impurity solver \cite{Werner06,Werner06b} which can treat the full rotationally invariant Hunds coupling. We focus mainly on the metallic regions of the phase diagram. Care must be taken to converge the solution (up to $30$ iterations of the dynamical mean field procedure are required for parameters near the metal-insulator transition line) and temperatures must be chosen low enough to reveal the quasiparticle behavior in all orbital sectors. 

In the single-site dynamical mean field approximation the metal-insulator transition is first order, characterized by an upper critical interaction strength $U_{c2}$ which marks the limit of stability of the metallic phase and a lower critical interaction strength $U_{c1}$ which marks the limit of stability of the insulating phase. Fig. ~\ref{phasedia} presents $U_{c1}$ of $\Delta$. The location of the rare earth nickelates on the phase diagram is not known. $LaNiO_3$ is metallic in bulk and (except for one and perhaps 2 monolayer samples) in thin film form. Other members of the family such as $NdNiO_3$ have insulating ground states, suggesting that the materials are close to a metal/charge-transfer insulator phase boundary, but  the origin of the insulating phase is controversial.\cite{Mazin07,Medarde09,Wang11} We therefore study a range of parameters in the metallic state. In DMFT the insulating state typically exhibits some form of orbital order, making the interpretation less clear. 

\begin{figure}[b]
   \includegraphics[width=0.95\columnwidth,angle=0.0]{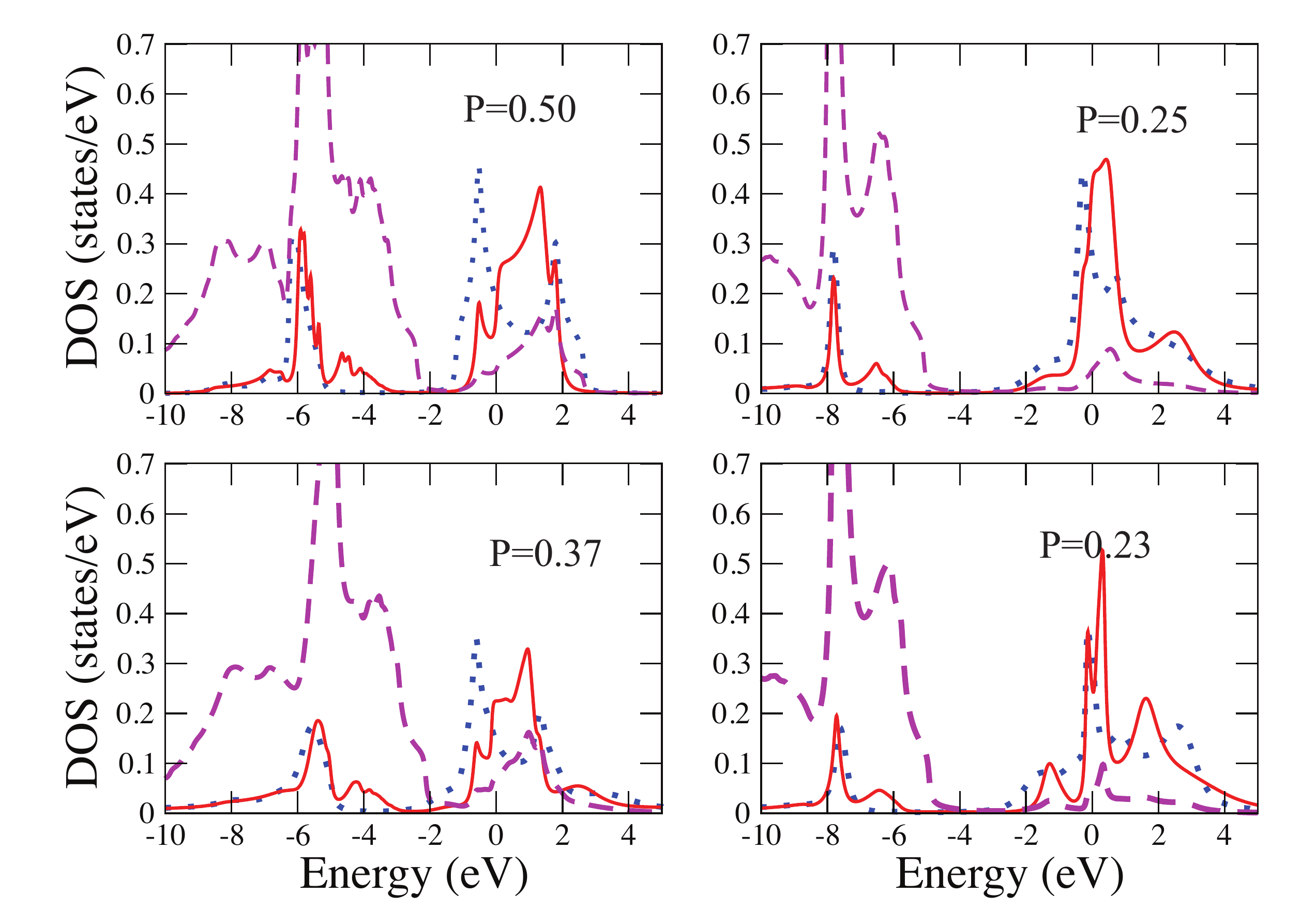}
   \caption{(Color online) Analytically continued spectral function $A(\omega)$, for
     Ni $d_{x^2-y^2}$ (dotted, blue on line), $d_{3z^2-r^2}$ (solid, red on-line) orbitals and sum of all O orbitals (dashed, magenta on line). Parameters ($N_d$ computed from integral of the  many-body DOS over the entire manifold of occupied states)
     (a) $U=0$, $N_d = 1.99$. $\varepsilon_{d}=-1.22$ eV,   $\varepsilon_p=-5.2$ eV.
     (b) $U=4$ eV, $N_d = 1.60$. $\varepsilon_{d}=-3.91$ eV, $\varepsilon_p=-7.89$ eV. 
     (c) $U=6$ eV, $N_d = 2.12$. $\varepsilon_{d}=-8.95$ eV, $\varepsilon_p=-4.93$ eV.
     (d) $U=6$ eV, $N_d = 1.52$. $\varepsilon_{d}=-5.75$ eV, $\varepsilon_p=-7.73$ eV.
     Fermi level is  zero. $J=0.5$ eV and $T=0.1$ eV. Computed polarization $P$ from Eq. ~[\ref{Pdef}] with $E_{low}=-3eV$ shown on  each panel. The $P$ values corresponding to integration over the energy range of the entire $p-d$ manifold are$P=(0.16,0.14,0.11,0.11)$ for panels a-d respectively.
     \label{dos}}\end{figure}

The many-body electronic structure is represented by  the local spectral function (many-body density of states)  $A_{a}(\omega)=Im\left(-i \int dt e^{-i\omega t}\left<\left[\psi_a(j,t),\psi^\dagger_a(j,t{'})\right]\right>\right)$ (here $a$ labels an orbital and $j$ a unit cell in the lattice).  Panel (a) of Fig.~\ref{dos} presents the noninteracting  ($U=0$) case, for which $\Delta\approx -4eV$. These spectra are  consistent with previously published GGA results\cite{Han10}  (the small differences arise from the difference between the  two dimensional model used here and the fully three dimensional model of Ref. ~\onlinecite{Han10}). Two energy regions are evident: a near Fermi surface region representing the $d-p$ antibonding band and a lower energy region representing the bonding combination of $d$ and $p$. The $p$ level energy is visible as a sharp peak in this energy region. Panel (b) shows the result of increasing the interaction strength to $U=4eV$ while keeping both the total electron count and  $\Delta$ fixed. Comparison to Fig.~\ref{phasedia} shows that this change moves the system close to the metal-insulator phase boundary. We see that the splitting between the bonding and antibonding regions of the spectrum increases, essentially because the interaction increases the effective d-level energy. We also see that the differences between the spectra of the two d orbitals are smaller than in panel (a). Panels (c) and (d) of Fig.~\ref{dos} shows spectra obtained for a stronger interaction $U=6eV$. Panel (c) shows  a double-counting correction corresponding to $\Delta=4eV$ and  chosen to undo the shift in the $d$-$p$ splitting. The parameters are far from the metal-insulator phase boundary and the spectra are seen to be very similar to those computed for the noninteracting model. In panel (d) we have chosen the double counting correction to keep the energy separation between the $O$ states and antibonding $p-d$ states approximately the same as in panel (b). The features near the Fermi level are narrower than in panel (b) because the system is closer to the metal-insulator phase boundary, but the spectra are otherwise similar. 

\begin{figure}[t]    \centering    \includegraphics[height=0.85\columnwidth, angle=0]{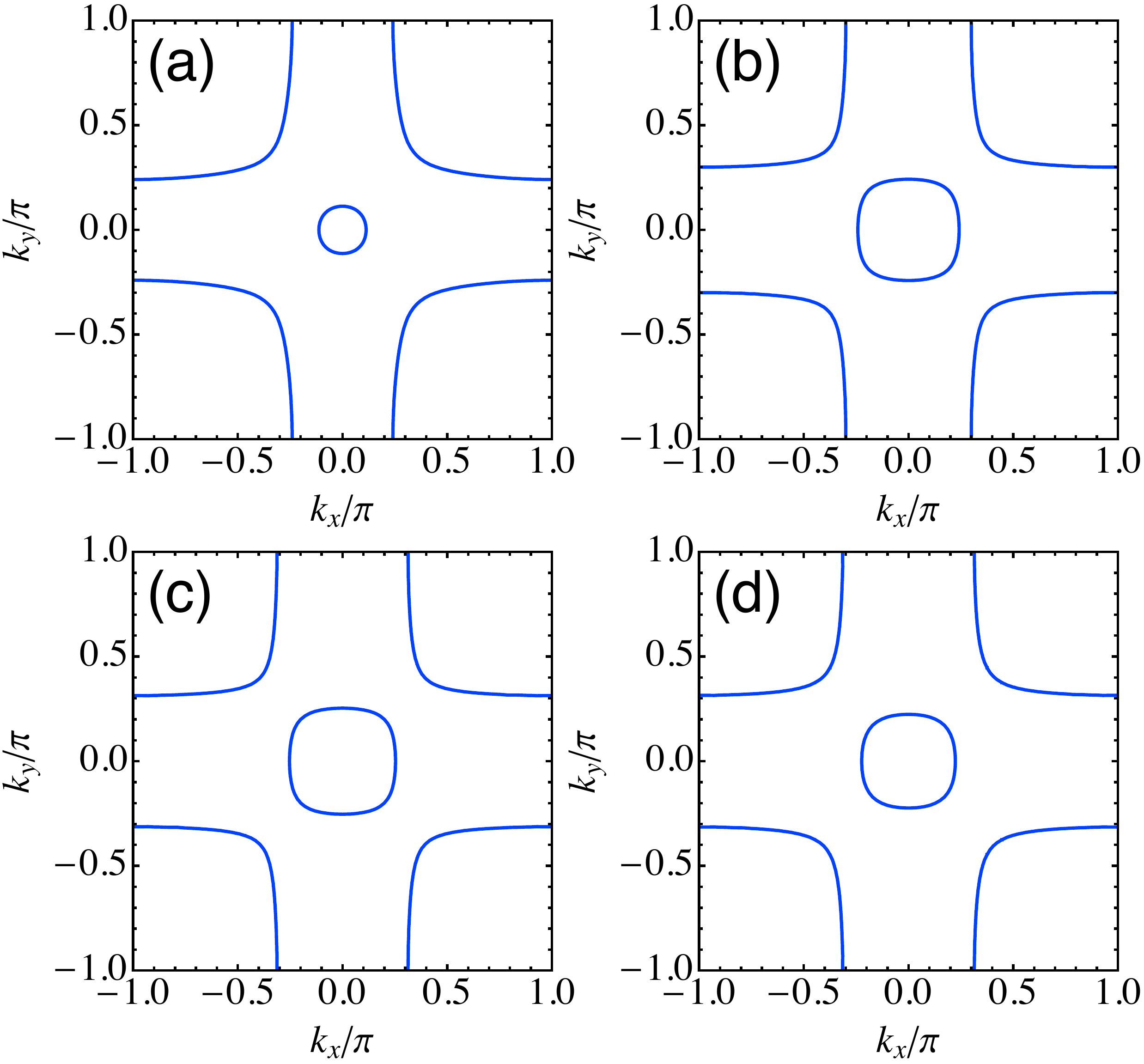}    \caption{Calculated Fermi surfaces at $J=0.5$ eV and  $\Delta=-1.98$ eV. (a) $U=0$, $N_d=2.44$, $\varepsilon_d=-1.81$ eV. (b) $U=4$ eV, $N_d=1.65$, $\varepsilon_d=-4.65$ eV.  (c) $U=6$ eV, $N_d=1.45$, $\varepsilon_d=-5.78$ eV. (d) $U=7$ eV, $N_d=1.34$, $\varepsilon_d=-6.02$ eV.  For panels (a)-(c) $T=0.05$ eV; for panel (d)  $T=0.025$ eV,}    \label{FSed}\end{figure}

The occupancy $n_a$  for a given orbital $a$ is defined as
\begin{equation}
n_a=\int_{E_{low}}^\mu \frac{d\omega}{\pi}A_a(\omega)
\label{nadef}
\end{equation} 
and the orbital polarization is 
\begin{equation}
P=\frac{n_{x^2-y^2}-n_{3z^2-r^2}}{n_{x^2-y^2}+n_{3z^2-r^2}}
\label{Pdef}
\end{equation}
We have chosen the zero of  energy such that $\mu=0$.   We believe it is most physically reasonable to focus on the difference in occupancy of near Fermi-surface states, corresponding to taking $E_{low}=-3eV$ to capture the antibonding but not the bonding bands. Alternatively, one may integrate over the whole (many-body) bandwidth. We have provided the P corresponding to both definitions; the results are very similar and the conclusions are not changed.   The computed $d$ occupancies  and polarizations are given in the caption and panels of Fig. ~\ref{dos}; interactions decrease the polarizations.

An alternative, low-energy definition of orbital polarization may be obtained from  the Fermi surface. In pseudocubic $LaNiO_3$ the calculated Fermi surface has two sheets, corresponding to the two relevant $d$ orbitals.\cite{Hamada93} In single-layer high-$T_c$ cuprates the Fermi surface has only one sheet, corresponding to a single relevant $d$ band, so one may identify a single-sheeted Fermi surface with an orbitally polarized low energy theory.  Fig. \ref{FSed} shows the evolution of the Fermi surface of the nickelate superlattice  as the interaction strength is increased at fixed $\Delta \approx -2$. The non-interacting model has a substantial degree of orbital polarization, as seen from the very small size of the central Fermi surface region,  but  as soon as the interaction is turned on the size of the central patch increases and then does not change over the entire metallic region, consistent with the values of $P$ given in the caption of Fig.~\ref{dos}. (The  Fermi surface in panel (d) is slightly smaller because at the lowest accessible temperature the fully coherent Fermi liquid state was not achieved.)

The small value of $P$ we find is  in agreement with recent resonant X-ray absorption experiments\cite{Freeland10}  but does not agree with results of previous dynamical mean field studies of  Hansmann and collaborators.\cite{Hansmann09,Hansmann10} While there are minor technical differences  (including the use, by Hansmann et. al. of an Ising approximation to the Hunds interaction) we believe that the most important issue is the model. Refs.~\onlinecite{Hansmann09,Hansmann10} downfolded the band theory results to a two-band model representing only the antibonding band, whereas in our work the $Ni$-$O$ charge transfer plays an important dynamical role, enabling the high spin $d^8{\bar L}$ configuration which is not susceptible to orbital polarization. 

\begin{figure}[t]
  \centering \includegraphics[width=0.85\columnwidth,angle=0]{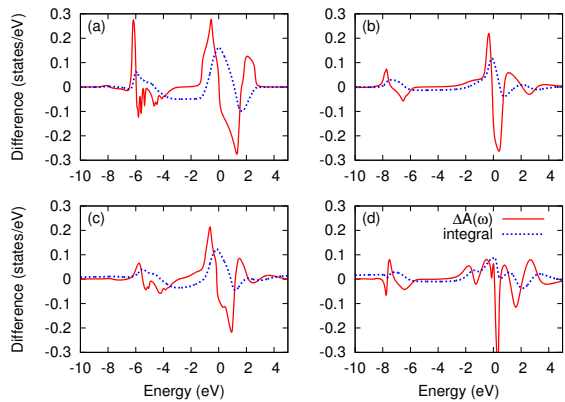}
   \caption{(Color online) d-spectrum difference plot 
     $\Delta A(\omega)=A_{x^2-y^2}(\omega)-A_{3z^2-r^2}(\omega)$ (solid line, red online). Parameters for (a)-(d) correspond to those in
     Fig.~\ref{dos}.
 \label{diff}}
\end{figure}

Our results suggest that in realistic models of nickelate superlattices, a significant orbital polarization will be very difficult to achieve. However, the two orbitals will not have identical properties. We present in Fig.~\ref{diff} the difference $\Delta A(\omega)=A_{x^2-y^2}(\omega)-A_{3z^2-r^2}(\omega)$ calculated for parameters corresponding to  Fig.~\ref{dos}(a)-(d). The d-spectral function may be measured in  resonant x-ray scattering experiments, and difference spectra are relatively insensitive to experimental complications such as core-exciton and final-state corrections. We see that the different electronic structures lead to observable effects on the spectra. In the N$_d \sim 2.0$ case (corresponding to panels (a) and (c) in Fig. ~\ref{dos}) the difference spectra reveal  prominent peaks just below the Fermi level ($\sim -0.5$ eV) and around $\sim +2.0$ eV. These features do not appear in the  $N_d \sim 1.45$ cases, (panels (b) and (d) in Fig~\ref{dos}). Also the two feature provide a measure of the physical $\varepsilon_p-\varepsilon_d$. Comparison of these calculations to new generations of X-ray absorption experiments\cite{Liu11,Freeland10} may help evaluate the orbital polarization and pin other material parameters.

In summary, we have shown that in a realistic many-body model of nickelate heterostructures, it is essentially not possible to achieve a significant degree of orbital polarization, so that the idea \cite{Chaloupka08} of obtaining a single-band electronic structure must be discarded. Further, we showed that a reduction of the full Hamiltonian to a Hubbard-like model \cite{Hansmann09,Hansmann10} which includes only the correlated orbitals yields a fundamentally misleading picture of the electronic structure. We presented spectra which should help in establishing the actual value of the double-counting correction for these materials, which is crucial to the metal-insulator transition behavior.

{\it Acknowledgements:}  AJM, MJH and CM were supported by the U. S. Army Research Office  via grant No. W911NF0910345 56032PH, XW by the Condensed Matter Theory Center. Part of this research was conducted at the Center for Nanophase Materials Sciences, sponsored at Oak Ridge National Laboratory by the Division of Scientific User Facilities, U.S. Department of Energy. The impurity solver is based on a code developed by P. Werner \cite{Werner06b} and uses the ALPS library \cite{ALPS}.

\bibliography{refs_superlattice}

\end{document}